\documentclass[rnote]{aa} 
\usepackage{graphicx}
\usepackage[varg]{txfonts}
\begin{document}
   \title{The Radio--X-ray Connection in Young Stellar Objects in the Orion Nebula Cluster}


   \author{Jan Forbrich
          \inst{1,2}
          \and
          Scott J. Wolk\inst{2}
          }

   \institute{Institute for Astrophysics, University of Vienna,
              T\"urkenschanzstra{\ss}e 17, 1180 Vienna, Austria\\
              \email{jan.forbrich@univie.ac.at}
         \and
	 Harvard-Smithsonian Center for Astrophysics, 60 Garden Street, Cambridge, MA 02138, USA\\
	      \email{swolk@cfa.harvard.edu}
             }

   \date{Received October 17, 2012; accepted --}

   \titlerunning{The Radio--X-ray Connection in YSOs}
   \authorrunning{Forbrich \& Wolk}

 
  \abstract
   {Both X-ray and radio observations offer insight into the high-energy processes of young stellar objects (YSOs). The observed thermal X-ray emission can be accompanied by both thermal and nonthermal radio emission. Due to variability, simultaneous X-ray and radio observations are a priori required, but only a comparably small number of YSOs have been studied in this way. Results have been inconclusive due to the even smaller number of YSOs that were simultaneously detected in X-ray and radio observations.}
   {We use archival X-ray and radio observations of the Orion Nebula Cluster (ONC) to significantly enlarge the sample size of known YSOs with both X-ray and radio detections.}
   {We study the ONC using multi-epoch non-simultaneous archival \textit{Chandra} X-ray and NRAO Very Large Array (VLA) single-band radio data. The multiple epochs allow us to reduce the impact of variability by obtaining approximated quiescent fluxes.}
   {We find that only a small fraction of the X-ray sources (7\%) have radio counterparts, even if 60\% of the radio sources have X-ray counterparts. YSOs with detections in both bands thus constitute a small minority of the cluster. The radio flux density is typically too low to distinguish thermal and nonthermal radio sources. Only a small fraction of the YSOs with detections in both bands are compatible with the empirical ``G\"udel-Benz'' (GB) relation. Most of the sources not compatible with the GB relation are proplyds, and thus likely thermal sources, but only a fraction of the proplyds is detected in both bands, such that the role of these sources is inconclusive.}
   {While the radio sources appear to be globally unrelated to the X-ray sources, the X-ray dataset clearly is much more sensitive than the radio data. We find tentative evidence that known non-thermal radio sources and saturated X-ray sources are indeed close to the empirical relation, even if skewed to higher radio luminosities, as they are expected to be. Most of the sources that are clearly incompatible with the empirical relation are proplyds which could instead plausibly be thermal radio sources. The newly expanded Jansky Very Large Array with its significantly enhanced continuum sensitivity is beginning to provide an ideal tool for addressing this issue. Combined X-ray and radio studies of YSOs using older VLA data are clearly limited by the typically low signal-to-noise of the radio detections, providing insufficient information to disentangle thermal and nonthermal sources.}


   \maketitle
%

\section{Introduction}

It has been know for some time that Young Stellar Objects (YSOs) produce observable X-ray and centimetric radio emission \citep[e.g., ][]{fei99} virtually throughout their various evolutionary stages, typically grouped into classes 0--III. Class 0/I protostars correspond to the earliest stages, followed by disk-dominated class II objects (classical T Tauri stars) to essentially diskless class III sources, also known as WTTS or weak-line T Tauri stars \citep{lad87,and93}.

The X-ray radiation from these objects is thought to be due to thermal emission from coronal-type activity. There is also non-thermal radio emission tracing magnetic fields that may be cospatial to, and confining, the X-ray emitting plasma. However, while such radio emission is observed as (gyro)synchrotron radiation, there is also thermal bremsstrahlung from ionized material in the disk or stellar winds which is strong at radio wavelengths (e.g., \citealp{dul85,and96,gue02}). 

Various types of active stars have been shown to have correlated X-ray and (nonthermal) radio luminosities according to the so-called G\"udel-Benz (GB) relation, $L_{\rm X}/L_{\rm R} \approx 10^{15\pm1}$~Hz extending over 10 orders of magnitude \citep{gue93,ben94}. Not strictly linear, the most luminous sources tend to be radio-overluminous. The correlation suggests common energy release and transformation processes in the coronae of active stars. For the most luminous subset, including the T Tauri stars, \citet{gue93} derive a proportionality constant of $L_{\rm X}/L_{\rm R} = 5.2\times10^{14}$~Hz, still with considerable scatter. The sources are all very active stars and thus likely have X-ray emission at saturation levels (e.g., \citealp{gue04}). Note that the most luminous sources considered by \citet{gue93} have radio luminosities of log($L_R$)=18 erg\,s$^{-1}$\,Hz$^{-1}$, the sources considered here are up to five times more luminous. The detailed interpretation of the empirical relation is not obvious, see \citet{gue02} and \citet{for11} for recent summaries. 

While the original results on the GB relation contained a handful of (class III) YSOs, these results were obtained from non-simultaneous observations and only the most luminous X-ray sources among YSOs could be detected at that time. An early discussion of the X-ray and radio properties of weak-line T Tauri stars can be found in \citet{whi92}. A series of experiments using simultaneous multi-wavelength observations has since been carried out to find out whether the GB relation holds for YSOs in general and for the earlier evolutionary stages in particular. Early studies targeted individual sources. \citet{fei94} observed V773~Tau and found that the variability of the nonthermal radio emission and the X-ray emission were decoupled. On the other hand, \citealp{bow03} observed a radio outburst of the Orion YSO GMR A that coincided with enhanced X-ray activity. Subsequently the combination of the \textit{Chandra} X-ray Observatory and the NRAO Very Large Array has been used to study several clustered star-forming regions in simultaneous X-ray and radio observations ($\rho$ Oph: \citealp{gag04}, CrA: \citealp{for07}, LkH$\alpha$\,101: \citealp{ost09}, IC 348 and NGC 1333: \citealp{fow11}). However, simultaneous X-ray and radio observations have so far been limited by the radio sensitivity. The above-mentioned surveys of five clusters have detected a combined total of only 16 YSOs that can be confidently placed on an $L_X$ vs. $L_R$ diagram (see Figure 2 in \citealp{fow11}). Most of these sources fall `below' the GB relation by up to two orders of magnitude (first discussed by \citealp{gag04}). However, it has remained unclear how representative these sources are. 

To considerably increase the sample size, we have abandoned the requirement that the radio and X-ray observations be carried out simultaneously. We study non-simultaneous but multi-epoch archival data of the Orion Nebula Cluster (ONC). The aim is to study how common X-ray and radio detections are among YSOs in a larger sample to constrain any underlying connections. By not requiring simultaneous observations, we cannot account for variability on short timescales, but we can still make assessments of time-averaged properties. Since there is multi-epoch information for most of the sources discussed here (with the exception of radio sources detected only found in an integration of all epochs), we can determine approximated quiescent flux levels and estimate the role of variability. We will see that, reassuringly, the Orion sources fall into the same part of the $L_X$ vs $L_R$ diagram as the YSOs from simultaneous observations. Previous results from simultaneous observations are thus important to guide the analysis of these data.

\section{Radio and X-ray observations of the Orion Nebula Cluster}

Extensive radio observations of the Orion Nebula Cluster were obtained in the 1990s with the NRAO Very Large Array (VLA) in its A configuration. At a wavelength of 3.5~cm, the resulting angular resolution is $\sim0\farcs3$. We use previously published results based on four epochs of Orion VLA data with two slightly different phase centers. The observations were carried out in the years 1994--1997; for more details, see \citet{zap04}.

In X-rays, the ONC has been the target of the Chandra Orion Ultra-deep Survey (COUP) which obtained a 838~ks exposure of the region over a continuous period of 13.2 days in 2003. A total of 1616 X-ray sources were identified and analyzed in depth. The project is described in detail in \citet{get05a} and references therein. It is important to point out that only $\sim10$\,\% of 1616 sources are likely extragalactic, the overwhelming majority being members of the Orion Nebula Cluster \citep{get05b}.

\section{Data analysis}

While the archival radio and X-ray data described here are non-simultaneous with several years in-between, they are more than snapshot observations. The X-ray observations span about ten days in 2005, and the radio data consist of four epochs obtained in the years 1994 to 1997.

As a result of their data reduction and analysis, \citet{zap04} report the detection of 77 radio sources in the ONC in two mostly overlapping primary beam areas of $5\farcm3$ width (FWHM). Of these, 54 sources were detected in individual epochs and 23 additional sources were found in the concatenated dataset combining all epochs. At the sensitivity of the combined dataset reported by \citet{zap04}, at a reported noise level of 0.03~mJy/beam, about five extragalactic sources can be expected in the FWHM primary beam area \citep{win93}.

The area covered by the VLA observations is considerably smaller than the area studied with \textit{Chandra} as part of the COUP project. Still, the combined primary beam areas (FWHM) of the VLA observations cover 623 X-ray sources reported by the COUP. Of these, 594 sources have enough S/N so that total X-ray luminosities corrected for foreground extinction could be reported.

The fact that both the radio and the X-ray observations consist of multiple epochs allows us to quantify the impact of variability like flaring on the averaged radio and X-ray luminosities that we derive. For the 54 radio sources detected in at least one of the four epochs, the median variability, measured as the factor between maximum and minimum flux density \footnote{One of these factors is a lower limit if the minimum flux is an upper limit (source 47).}, is a factor of 2.2. The maximum variability seen is a factor of 43 -- it is GMR A. This, however, is exceptional, as only three sources out of 54 are variable by more than a factor of 10, the other two sources being numbers 33 and 77 in \citet{zap04}. For the COUP X-ray data, \citet{get05a} estimate that 60\% of the sources are variable, often by an order of magnitude. Since both the radio and X-ray luminosity estimates contain flares, they will on average overestimate the ``quiescent'' (but hard to quantify) X-ray and radio luminosities, but typically not by more than an order of magnitude.

To not be biased by individual large flares in either the radio or X-ray ranges, we use only the \textit{minimum} reported luminosities to approximate the quiescent fluxes. In the radio range, we use the minimum radio flux density reported by \citet{zap04} for sources detected in individual epochs (or the flux density from the integration of all epochs for the weakest sources). To not lose the few sources with multiple detections but occasional nondetections, we use the lowest reported flux or the lowest upper limit, whatever is lower. In the X-ray range, we scale the reported absorption-corrected averaged X-ray luminosity by the ratio of the minimum and average total count rate, as reported in the Bayesian block analysis by \citet{get05a}. This gives us a measure of the minimum X-ray luminosity. 

To cross-match the radio and X-ray datasets, we use a matching radius of $0\farcs3$ which corresponds to the peak of the histogram of source match separations and also the approximate angular resolution of the VLA data. Within this search radius, we identify 46 sources as both X-ray and radio emitters. Only two of these sources are listed as possibly extragalactic by \citet{get05b}. 60\% of the reported radio sources have X-ray counterparts, but only 7\% of the reported X-ray sources are detected in the radio band. Increasing the search radius to $0\farcs5$ yields only five more sources, but already includes one match to two sources. A complication comes from the fact that compact radio emission of YSOs may be unobservable if these are behind an optically thick screen of ionized material in the foreground. Such material may not be observable in the present data since extended structure is spatially filtered out in the interferometric observations.

However, a potential explanation for this difference could also be a sensitivity difference between the X-ray and radio datasets. We could be detecting only the brightest part of the radio population and the more sensitive X-ray dataset would contain many sources with radio emission that is too weak. Such an effect would become clear in radio and X-ray luminosity histograms of the sample. We first convert all luminosities to a common distance of 414~pc \citep{men07}.  While it is difficult to quantify the relative sensitivity of the two datasets, the dynamic range of the two datasets is very different: Already in our sample of sources with X-ray and radio detections, the radio flux densities span two orders of magnitude, while the X-ray fluxes span four orders of magnitude. Taking into account all sources in the field of view, the X-ray sources cover six and the radio luminosities three orders of magnitude in luminosity. Assuming similar luminosity distributions, this suggests that the X-ray dataset is considerably more sensitive than the radio dataset.

\begin{figure*}
\begin{minipage}{0.5\linewidth}
\includegraphics[angle=-90,width=1.05\linewidth]{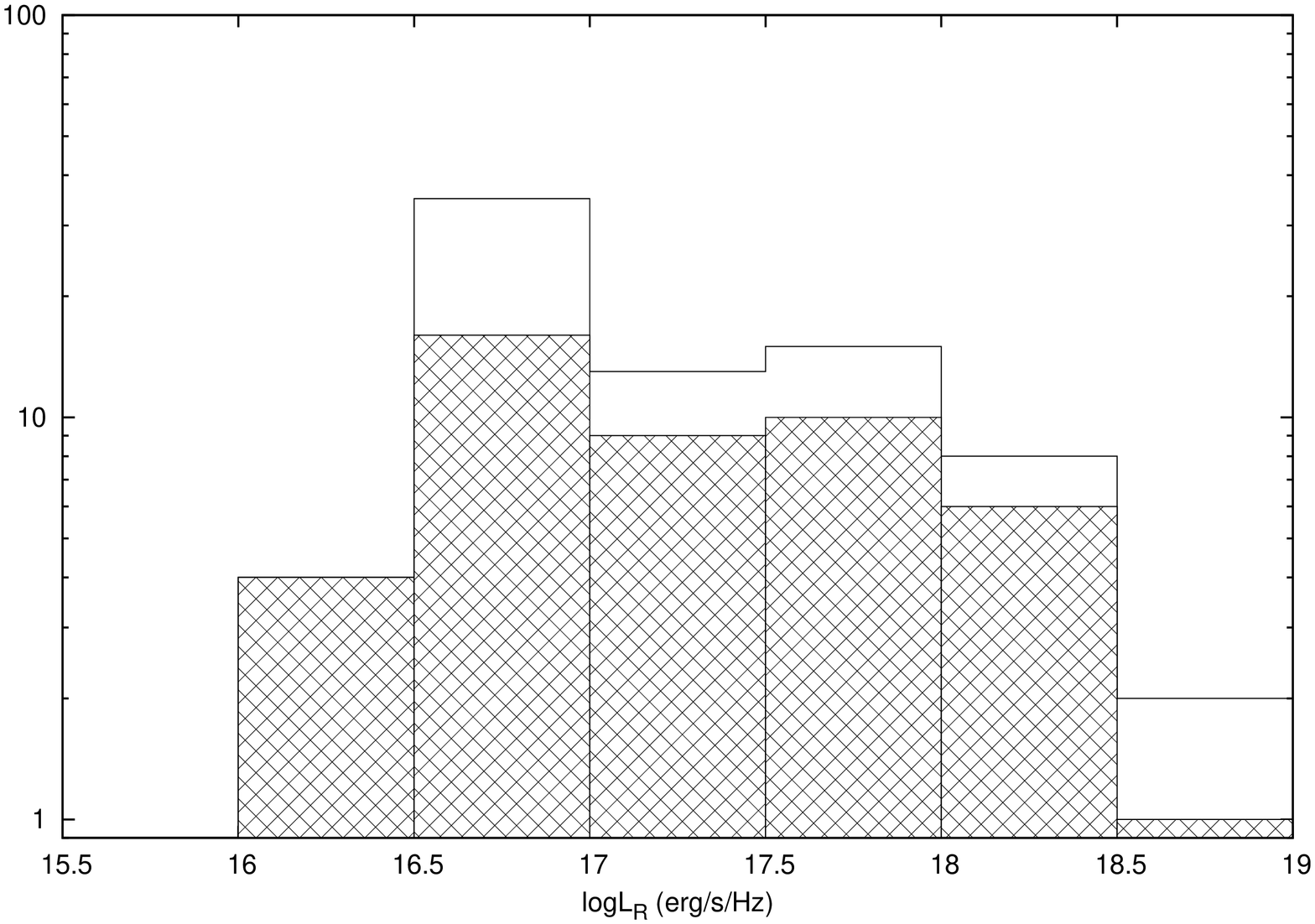}
\end{minipage}
\begin{minipage}{0.5\linewidth}
\includegraphics[angle=-90,width=1.05\linewidth]{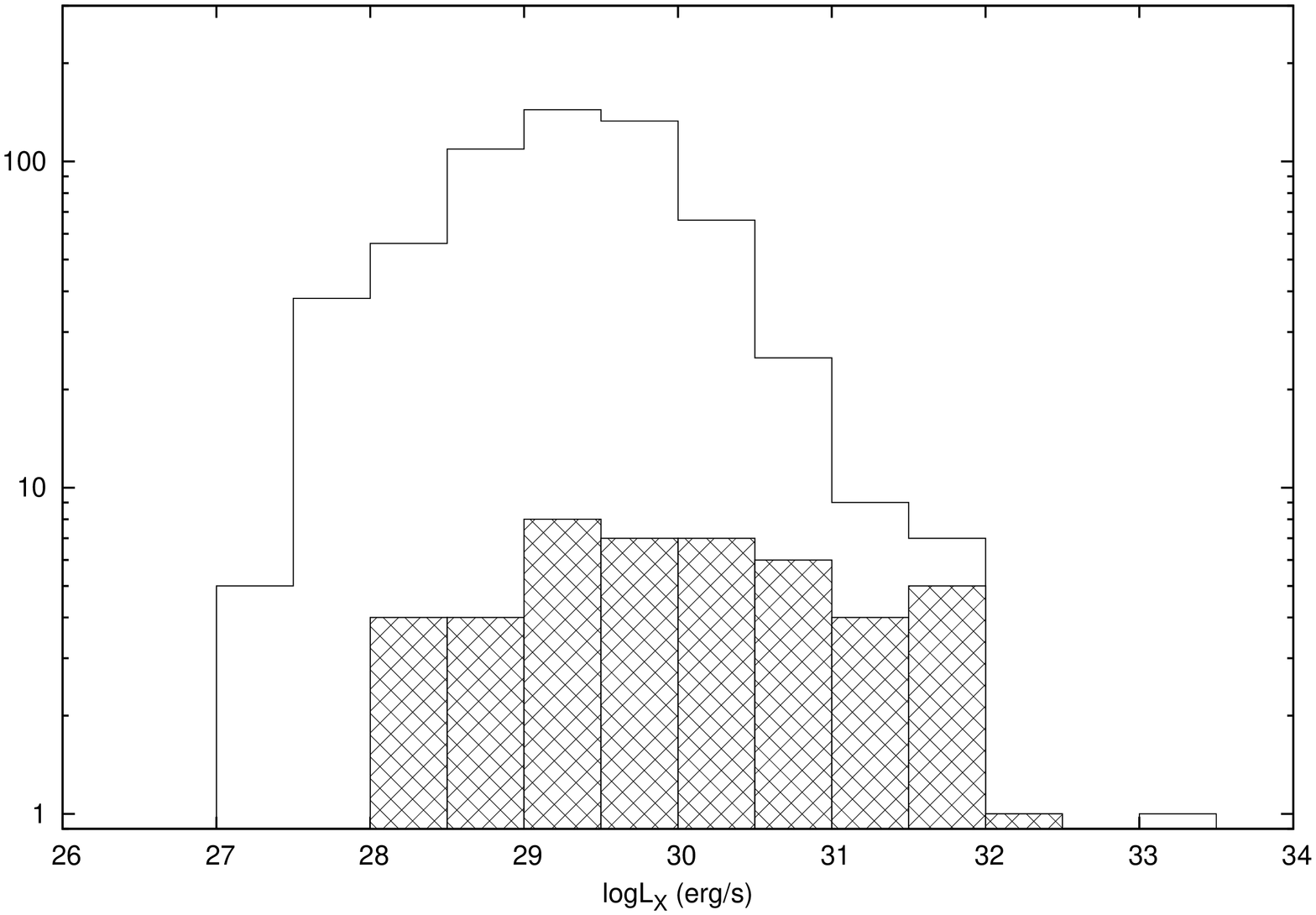}
\end{minipage}
\caption{Left panel: Histogram of the radio luminosities of all sources detected by \citet{zap04} (empty histogram) and those with X-ray counterparts (hashed histogram). Right panel: Histogram of the COUP X-ray luminosities in the area probed by \citet{zap04} with the VLA (empty histogram) and those with radio counterparts (hashed histogram) on a logarithmic scale.\label{fig_histRX}}
\end{figure*}

In Fig~\ref{fig_histRX}, we show the distribution of the quiescent radio and X-ray luminosities of our sample, derived as described above. The left hand panel displays the distribution of radio luminosities, both for the entire sample and for the X-ray detected subsample. The fraction of radio sources with X-ray counterparts rises toward the highest radio luminosities, from 50\% to a maximum of 80\%. A histogram of the X-ray luminosities is shown in the right hand panel of Fig~\ref{fig_histRX}. While the most luminous X-ray sources have radio counterparts, the detection fraction quickly drops to below $\sim10$\,\% for lower X-ray luminosities. Corroborating the picture of the more sensitive X-ray dataset, the weakest radio sources still have X-ray counterparts while that is not the case vice versa. Note that the most luminous X-ray sources here are young high-mass stars, not low-mass YSOs, and energetic winds are understood to play an important role in their X-ray and radio emission (e.g., \citealp{gue09}).

\begin{figure*}
\centering
\includegraphics[bb =100 55 352 300, width=\linewidth]{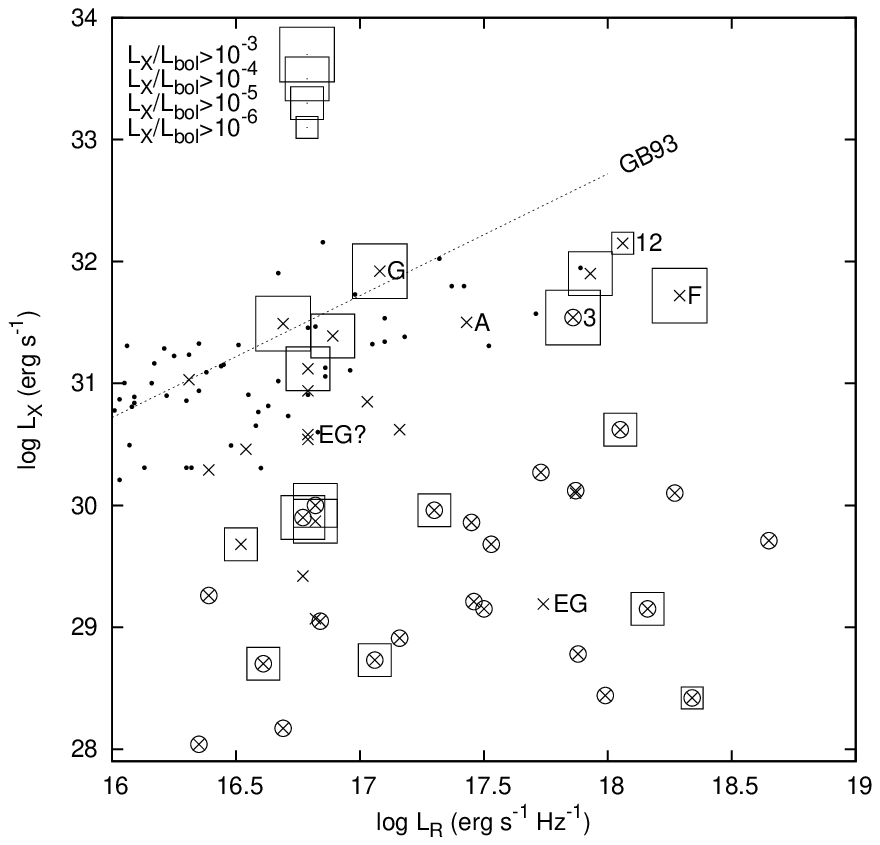}
\caption{$L_X$ vs. $L_R$ diagram of the ONC, based on COUP X-ray data and radio observations reported by \citet{zap04}, detections are marked by `$\times$'. The dotted line marked `GB93' marks the relation reported for WTTS (among other sources) by \citet{gue93}, the filled dots mark all sources in \citet{gue93} that fall into the luminosity range considered here. The squares mark sources with $L_X/L_{\rm bol}$ information, as explained in the upper left corner. Proplyds from \citet{kas05} are marked by `$\circ$'. Some sources are marked with their GMR names (see text) and the only two (possibly) extragalactic sources from \citet{get05a} are marked by `EG'.\label{fig_lxlr}}
\end{figure*}

A direct comparison of the present dataset to the GB relation can be obtained by plotting the X-ray versus the radio luminosities. In Figure~\ref{fig_lxlr}, we show such a plot together with the expectation from the GB relation. While by far not all X-ray sources have radio counterparts and vice versa, we here only show the 46 sources with both X-ray and radio detections. As detailed below, upper limits for nondetections would be potentially misleading if due to intervening foreground material. The plot shows that the radio--X-ray detections are not scattered around the GB relation, instead they are shifted toward higher radio (or lower X-ray) luminosities, and they do not form a tight correlation. The detections lie up to almost five orders of magnitude below the GB relation. While one has to keep in mind that the dataset discussed here consists of non-simultaneous observations, we have taken care to select only the minimum, presumably approximately quiescent fluxes. Also, as discussed above, variability is estimated to typically account for less than an order of magnitude in flux and thus cannot explain this discrepancy. In the following, we discuss three distinct groups of sources.

\subsection{Sources with both X-ray and radio detections}

Only 46 out of 623 X-ray sources (and 77 radio sources) have detections in both X-ray and radio emission. All but one of these sources fall `below' the GB relation, some by almost five orders of magnitude. Judging strictly by the empirical relation, most X-ray sources should have radio counterparts below the detection limit of the present data. It is in this sense that the sources with detections in both bands are `radio-bright'.

As mentioned above, the GB relation is expected to hold only for nonthermal gyrosynchrotron radio sources. Since for most of the radio sources the emission mechanism is unknown, we check VLBI centimeter continuum detections of Orion YSOs as unambiguous signs of nonthermal emission. \citet{men07} report VLBI detections of sources GMR A, 12, G, and F (see also \citealp{fel89,fel91} for VLBI observations of GMR 12 and \citealp{san07} for VLBI detections of GMR A). Note that \citet{zap04} find circular polarization, indicative of gyrosynchrotron radiation, only toward GMR A (the flare source from \citealp{bow03}). These nonthermal radio YSOs are indeed close to, but still below the GB relation for WTTS.

We can also check for dependencies on spectral type. The COUP database provides spectral types for 24 out of the 46 sources with both X-ray and radio counterparts. The earliest spectral type among the sources detected in both bands is GMR 12, listed as O9--B2. We have seen that GMR 12 is, in fact, a nonthermal radio source (see above), but the presence of tight binaries cannot be ruled out in this sample. With the exception of a single F star, the other reported spectral types are all G, K, and M. In the available data, there is no apparent correlation of spectral type and the location in the $L_X$ vs. $L_R$ diagram of sources with detections in both bands.  

The information provided by the COUP database also allows us to address whether there are saturated X-ray sources in our sample and where they are on the $L_X$ vs. $L_R$ plot. Out of the 46 sources with both X-ray and radio counterparts, 18 have listed bolometric luminosities. Ten of these are either saturated or close to saturation -- four at $\rm{log}(L_X/L_{\rm bol})>-3$ and six at $\rm{log}(L_X/L_{\rm bol})>-4$, even with the approximately quiescent luminosities derived above. For a discussion of the range in the $L_X/L_{\rm bol}$ ratios of the COUP sources, see \citet{pre05ea}. 
It should be pointed out that the four sources with $\rm{log}(L_X/L_{\rm bol})>-3$ have listed masses of 2--3 $M_\odot$ and are thus in the high-mass range in the context of \citet{pre05ea}, as are two of the six sources with $\rm{log}(L_X/L_{\rm bol})>-4$. The sources with $L_X/L_{\rm bol}$ information are marked in Figure~\ref{fig_lxlr}. Most of the X-ray sources with high X-ray luminosities and radio counterparts are indeed saturated. Sources with higher $L_X/L_{\rm bol}$ ratios clearly are closer to the GB relation than those with lower ratios. A clear exception is source 12, but as noted above, this is a high-mass source, potentially an O star, that cannot be directly compared to the lower-mass sources. Given the potential presence of multiple systems, the interpretation of the $\rm{log}(L_X/L_{\rm bol})$ ratio is not straightforward for sources of a few solar masses. The low-mass sources with $-4<\rm{log}(L_X/L_{\rm bol})<-3$ are below the GB relation by a factor of up to 40. Thus, while the data presented here include some X-ray--saturated sources, this effect contributes to understanding the upper bound of the X-ray luminosities, not their wide range.

While there is no expectation that the radio emission would depend on X-ray emission parameters, we can finally check for a dependence on absorbing column densities, X-ray temperatures, or mean X-ray photon energies (as listed in \citealp{get05a}). We find that when compared to the full sample, the X-ray sources with radio counterparts do not seem to be different in terms of these parameters (not shown).

\subsection{X-ray sources with radio upper limits}
92.6\% of the X-ray sources do not have radio counterparts. However, judging by the GB relation and an assumed range of plus and minus one order of magnitude, all but the most X-ray luminous of these sources, those with log$L_X>$30.7 erg\,s$^{-1}$, could have counterparts with radio fluxes below the detection limit. Thus, all of these objects may actually be compatible with the GB relation, some at considerably lower radio luminosities. Also, any radio emission of these sources could be extinguished by foreground ionized material. They therefore do not contribute much to our discussion of whether we find sources that do not follow the GB relation. We also know of one radio source that is undetected in the present multi-epoch radio dataset due to variability. It is the flare source ORBS from \citet{for08} which was detected at 8.5 GHz \citep{men95} in an observation epoch that was not included in the study by \citet{zap04}. This source has an X-ray counterpart in the COUP survey from which it was deduced that it is a deeply embedded YSO. So far, it has only been detected in the X-ray and radio wavelength regimes.

\subsection{Radio sources with X-ray upper limits}
The 40.3\% of radio sources without X-ray counterparts are more interesting since their X-ray upper limits lie several orders of magnitude below the GB relation. Only one of the 31 radio sources without X-ray counterparts has a 2MASS counterpart within $0\farcs5$, i.e., if these sources are ONC members, they are very deeply embedded. Searching for counterparts with the SIMBAD database yields several YSOs, including, most prominently, the BN object and source I. Also, 15 sources are within $0\farcs5$ of previously known YSOs. They are listed as ionized disks seen in emission by \citet{rrs08}. The X-ray nondetections of these sources could be due to variability or very high absorption. As an example of the former, consider again the case of the above-mentioned source ORBS \citep{for08} which was only detected in the COUP survey in a bright flare shortly before the end of the long observation.

\section{Discussion and Conclusions}

We have obtained a large consistent sample of YSOs with both radio and X-ray detections from archival data of the ONC. The multi-epoch nature of both datasets allows us to approximate quiescent flux levels in both bands by using the minimum fluxes at which the sources have been detected. We corroborate earlier results that the radio detections lie significantly `below' the GB relation in an $L_X$ vs. $L_R$ diagram -- by up to almost five orders of magnitude. The sample also shows a low percentage of sources with both X-ray \textit{and} radio detections. However, the number of such sources has increased compared to previous results from simultaneous observations. Comparisons of the detection fractions, while very different from cluster to cluster (including CrA with almost 100\%), are severely limited by the small sample sizes. The ONC now has well-determined detection fractions. The results allow us to quantify the connection, or rather the disconnect, between X-ray and radio emission in the ONC, with only 7\% of the X-ray sources also detected in the radio range. The location of these sources in the $L_X$ vs. $L_R$ plot does not seem to correlate with spectral type or X-ray emission parameters (absorbing column densities, X-ray temperatures, and mean photon energies). The main shortcoming of the radio data used here is the absence of information on the emission mechanism. However, there is tentative evidence that nonthermal radio sources and X-ray--saturated sources are relatively close to the GB relation, even if skewed to higher radio luminosities when compared to the relation reported for TTS by \citet{gue93}.

\paragraph{Proplyds} One might expect that the numerous proplyds detected in the ONC could have distinct X-ray and radio properties. \citet{kas05} analyzed the X-ray properties of 150 proplyds that have been identified in the ONC. They report an X-ray detection rate of about 70\%. This is only slightly lower than the general detection rate which is probably about 90\%, missing both absorbed and very low mass YSOs (Megeath, private comm.). The proplyds do not seem to account for any of the radio populations discussed above. There are 109 proplyds in the VLA field of view and while 76 of them have X-ray counterparts (i.e., 70\%), about a third of them, 37, have radio counterparts within $0\farcs3$ (the selection radius corresponds to the peak of the match per search radius histogram). Out of the latter 37 sources, 24 also have X-ray counterparts. At 32\%, the radio detection fraction of the proplyds with X-ray counterparts thus clearly exceeds the average detection rate of 7\%. Note that 20 sources have neither X-ray nor radio detections. However, all 37 sources with radio counterparts lie considerably below the GB relation (see Figure~\ref{fig_lxlr}), and in fact constitute most of the sources in this part of the plot. This could be due to the presence of ionized material in these systems causing thermal radio emission; proplyds would then not be expected to follow the GB relation. On the other hand, whether the proplyds play an important part of the explanation of the overall discrepancy remains unclear since their overall detection fraction at both X-ray and radio wavelengths is only 22\%. 

The sources that lie far off the expected relation could thus have a radio emission mechanism that is different from the nonthermal sources that are relatively close to the GB relation; most likely, that would be thermal emission that can be produced in jets and at the base of outflows, for example. Since this thermal emission would not be due to the central source, this would also explain why there is no apparent correlation between the radio emission and parameters of the central star or its X-ray emission. It is also important to keep in mind the large number of sources that are only detected in either radio or X-ray emission, but not in both. Since circumstellar dust is virtually transparent to centimetric radio emission, radio observations also trace sources that are so deeply embedded that any X-ray emission would no longer be observable. The radio detections without X-ray counterparts could be in this category. On the other hand, radio emission can be concealed by optically thick ionized gas along the line of sight, either in the immediate vicinity of the star or if the line of sight passes ionized portions of the nebulosity.

Currently, the assessment of the X-ray--radio connection in YSOs is clearly limited by the radio sensitivity. The present radio dataset of Orion YSOs lacks the sensitivity to obtain meaningful statements on polarization as a means of identification of the emission mechanism and it has no information on spectral indices. Also, with its lack of time resolution on time scales of hours or even minutes, the radio data discussed here cannot help in the assessment of the connection between X-ray and radio variability, particularly in flares. This deficiency in the radio data can soon be overcome, however, as the expanded capabilities of the Karl G. Jansky VLA (JVLA) are becoming available. 

\begin{acknowledgements}
We would like to thank the referee of this paper, Manuel G\"udel, for constructive comments that helped to significantly improve this paper. This publication is supported by the Austrian Science Fund (FWF). The National Radio Astronomy Observatory is a facility of the National Science Foundation operated under cooperative agreement by Associated Universities, Inc.
\end{acknowledgements}

\bibliography{bib_orionXR} 

\end{document}